\documentclass[%
reprint,
superscriptaddress,
showpacs,
amsmath,amssymb,
aps,
prl,
floatfix
]{revtex4-1}

\usepackage{graphicx}
\usepackage{dcolumn}
\usepackage{bm}

\usepackage{graphicx}

\begin{document}

\title{Near-field radiative heat transfer between macroscopic planar
surfaces}

\author{Richard Ottens} \affiliation{Department of Physics,
University of Florida, P.O. Box 118440, Gainesville, FL 32611-8440,
USA}

\author{V. Quetschke} \affiliation{Department of Physics and
Astronomy, University of Texas at Brownsville, 80 Fort Brown,
Brownsville, TX 78520, USA}

\author{Stacy Wise} \altaffiliation{Current address: Department of
Physics and Atmospheric Science, Dalhousie University, Halifax, NS 
B3H3J5, Canada} \affiliation{Department of Physics, University of
Florida, P.O. Box 118440, Gainesville, FL 32611-8440, USA}

\author{A.A. Alemi} \altaffiliation{Current address: Department of
Physics, Cornell University, Ithaca, NY 14853-2501, USA}
\affiliation{Department of Physics, University of Florida, P.O. Box
118440, Gainesville, FL 32611-8440, USA}

\author{R. Lundock} \altaffiliation{Current address: Astronomical
Institute, Tohoku University, Aoba, Sendai 980-8578, Japan}
\affiliation{Department of Physics, University of Florida, P.O. Box
118440, Gainesville, FL 32611-8440, USA}

\author{G. Mueller} \affiliation{Department of Physics, University
of Florida, P.O. Box 118440, Gainesville, FL 32611-8440, USA}

\author{D.H. Reitze} \affiliation{Department of Physics, University
of Florida, P.O. Box 118440, Gainesville, FL 32611-8440, USA}

\author{D.B. Tanner} \affiliation{Department of Physics, University
of Florida, P.O. Box 118440, Gainesville, FL 32611-8440, USA}

\author{B.F. Whiting} \affiliation{Department of Physics, University
of Florida, P.O. Box 118440, Gainesville, FL 32611-8440, USA}

\date{\today}

\begin{abstract}

Near-field radiative heat transfer allows heat to propagate across a
small vacuum gap in quantities that are several orders of magnitude
greater then the heat transfer by far-field, blackbody radiation.
Although heat transfer via near-field effects has been discussed for
many years, experimental verification of this theory has been very
limited. We have measured the heat transfer between two macroscopic sapphire
plates, finding an increase in agreement with expectations from
theory. These experiments, conducted near 300 K, have measured the
heat transfer as a function of separation over mm to $\mu$m and as a
function of temperature differences between 2.5 and 30 K. The 
experiments demonstrate that evanescence can be put to work to 
transfer heat from an object without actually touching it.

\end{abstract}


\maketitle

Humans knew of radiative heat transfer at least as early as the
discovery of fire, and physicists have investigated this process for
centuries, culminating in the black-body theory of Planck and the
birth of the quantum theory. Planck's equation for black-body
radiation contains only the temperature and some fundamental
constants.  When actual materials are involved, their emissivities
enter the discussion, but little else. For example, the heat
transfer per unit area between two semi-infinite planes is set by
their temperatures and integrated emissivities but does not depend
on their separation or other geometrical quantities. When the two
planes approach each other closely the situation changes. In this
near-field regime, each material interacts with exponentially
decaying evanescent electromagnetic fields generated in and existing
outside the other material; these fields can drive currents and
generate heat.\cite{Rytov,Polder,Maris} This near-field radiative
heat transfer can be several orders of magnitude greater than
far-field blackbody radiation.

Much like the Casimir and van der Waals force, near-field heat
transfer deals with fluctuations that only exist over small
distances. The first in-depth theory for near-field heat transfer
between planar surfaces was derived by Polder and Van
Hove,\cite{Polder} building on the work of Rytov\cite{Rytov}. There
have been several other theoretical approaches, and in general the
theory seems complete, except perhaps at distances comparable to
atomic dimensions.\cite{Kittel}

Although heat transfer via near-field effects has been discussed for
many years, experimental verification of the theory for heat
transfer between two planar surfaces has been limited.
Hargreaves\cite{Hargreaves} has presented room temperature
observations for two Cr surfaces at distances as small as 1~$\mu$m.
Domoto et al.\cite{Domoto} reported results at cryogenic
temperatures but for relatively large (50 $\mu$m) separations, where
near-field effects were barely observable. Neither study compared
experiment to theory. A comparison at a fixed spacing has been put
forward, but the plates were separated by polyethylene spacers, so
the distance could not be varied.\cite{Narayanaswamy} There have
also been several recent results using a sphere-plane
geometry.\cite{Kittel,Narayanaswamy,Chevrier,Shen} There are
engineering reasons for this approach, as, unlike a parallel plane
geometry, a sphere-plane geometry needs no angular alignment. Using
scanning probe and micro-machine technologies, these experiments
cover a wide distance range.

In this paper, we report a set of measurements of heat transfer
between two planar surfaces and present a detailed comparison to
theory. In addition to its intrinsic interest, this work was
motivated by possible applications in cooling objects without
actually touching them, such as the mirrors of a future  laser
interferometer gravitational-wave detector.\cite{Wise} This
application would require the parallel plane geometry in order to
get large areas of close approach of the two objects and thus
significant heat transfer.

The near-field heat-transfer process can be thought of as a form of
frustrated total internal reflection. Evanescent waves,
exponentially-decaying electromagnetic fields, exist outside, but
near to, the surface of a material medium at temperature $T$. 
These decaying fields
are a consequence of travelling waves inside the medium experiencing
total internal reflection; this phenomenon occurs when there is no
valid solution to Snell's Law, $n_{i} \sin \theta_{i} = n_{t} \sin
\theta_{t}$, where $n_{i}$ and $n_{t}$ are the indices of refraction
and $\theta_{i}$ and $\theta_{t}$ the angle between wave vector and
surface normal on the two sides of the interface. Although there is
no energy transmission across the interface, an electric field
exists on the far side. Furthermore, if another medium is brought
near this exponentially decaying field some of the energy from the
incident beam will propagate across the gap and into the new
material. If one medium is hotter than the other, this photon
tunneling will lead to heat transfer from hot to cold.

Polder and Van Hove\cite{Polder} considered two half spaces of
identical material but different temperatures. The computed heat
transfer coefficient $\mathcal{W}$ 
comes from the temperature derivative of the $z$ component of the
Poynting vector ($S_{z}$) from each medium, which is evaluated just inside 
surface of the
other medium.
 $\mathcal{W}$ is the temperature derivative of this difference as can be seen in Eq.~\ref{EQ:W}.
\begin{equation}
    \label{EQ:W}
  {\mathcal{W}}=\lim_{T_{1}-T_{2} \rightarrow 0} |\frac{S_{z}(d_{+}) - S_{z}(0_{-})}{T_{1}-T_{2}}| = \frac{\partial S_{z}(d_{+})}{\partial T_{1}}
\end{equation}

It is convenient to break $\mathcal{W}$ into two components
\begin{equation}
    \label{EQ:W2}
    \mathcal{W} = \mathcal{W}_{\rm sin} + \mathcal{W}_{\rm exp}.
\end{equation}
$\mathcal{W}_{\rm sin}$ is the part where the wave number $k_{x}$ of the field
is in the range $0<k_{x}<\omega/c$, where the field is propagating,
and which gives the ordinary (Stefan-Boltzmann) far-field radiation.
$\mathcal{W}_{\rm exp}$ is the component where the field is
exponentially decaying away from the surface; it is the larger
contributor to the near-field
limit:
\begin{equation}
    \label{EQ:Wexp}
    \mathcal{W}_{\rm exp} = \int^{\infty}_{0} d\omega \int^{\infty}_{\omega/c} dk_{x} \frac{k_{x}}{4\pi^{2}} 
(\mathcal{T}^{\rm exp}_{\|} + \mathcal{T}^{\rm exp}_{\bot}) \frac{\partial [\frac{\hbar\omega}{e^{\hbar\omega/{kT}}-1}]}{\partial T}
\end{equation}
where the energy transmission coefficients are 
\begin{equation}
    \label{EQ:tepar}
    \mathcal{T}^{\rm exp}_{P} = \frac{1-\cos (2\chi_{P})} {\cosh[2 \kappa 
(d-\delta_{P})]-\cos (2\chi_{P})},
\end{equation}
with $P = \|$ or ${\bot}$ for the two polarizations and 
$\chi_{\|}$ and $\chi_{\bot}$ being the phase shifts on reflection
\begin{equation}
    \label{EQ:Xpar}
    \chi_{\|} = \arg[(-i \kappa \epsilon + k_{z})(-i \kappa \epsilon - k_{z})^{*}],
\end{equation}
\begin{equation}
    \label{EQ:Xper}
    \chi_{\bot} = \arg[(-i \kappa + k_{z})(-i \kappa - k_{z})^{*}],
\end{equation}
with $\epsilon$ the material dielectric function. $\delta_{\|}$ and $\delta_{\bot}$ can be derived from
\begin{equation}
    \label{EQ:dpar}
    e^{2 \kappa \delta_{\|}} = \left|\frac{i \kappa \epsilon + k_{z}}{i \kappa \epsilon - 
k_{z}}\right|^{2},
\end{equation}
\begin{equation}
    \label{EQ:dper}
    e^{2 \kappa \delta_{\bot}} = \left|\frac{i \kappa + k_{z}}{i \kappa - 
k_{z}}\right|^{2},
\end{equation}
and $k_{z}$ and $\kappa$ are the $z$ component of $k$ for the medium and the vacuum, respectively
\begin{equation}
    \label{EQ:Ksz2}
    k_{z} = \sqrt{(\epsilon - 1) \omega^{2} / c^{2} - \kappa^{2}},
\end{equation}
\begin{equation}
    \label{EQ:Ksvz2}
    \kappa =  i k_{zv} = \sqrt{k^{2}_{x} - \omega^{2} / c^{2}}.
\end{equation}
Note that  $k_{x} > \omega/c$, so that $k_{zv}$ is imaginary above the surface and the energy transmission coefficients 
$\mathcal{T}^{\rm exp}_{\|}$ and $\mathcal{T}^{\rm exp}_{\bot}$ are not  derived from the energy reflection coefficients $\mathcal{R}_{\|}$ and 
$\mathcal{R}_{\bot}$.\cite{Polder}
We have used
this theory to calculate the heat transfer coefficient between
sapphire plates, using Barker's sapphire dielectric function.\cite{Barker}

\begin{figure}
\includegraphics[width=.8\columnwidth]{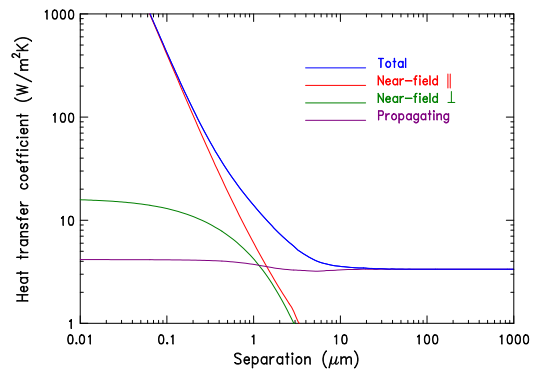}
\caption{\label{Fig:WforSapphire} Heat transfer coefficient
vs.~distance for z-cut sapphire. The temperatures of the two media
are $T_{\rm hot}=310$ K and $T_{\rm cold}=300$ K, respectively. The
violet curve shows $\mathcal{W}_{\rm sin}^{\|} + W_{\rm sin}^{\bot}$
which dominate at far distances. The red curve shows
$\mathcal{W}_{\rm exp}^{\|}$. The green curve shows
$\mathcal{W}_{\rm exp}^{\bot}$. The evanescent terms dominate at
close distances. The blue curve is the total heat transfer
coefficient
$\mathcal{W}$.}
\end{figure}

Figure \ref{Fig:WforSapphire} shows the prediction of the model for
z-cut sapphire; the temperatures of the two media are $T_{\rm
hot}=310$ K and $T_{\rm cold}=300$ K. The $\mathcal{W}_{\rm sin}$
term dominates at large separation and is nearly constant with
separation. The $\mathcal{W}_{\rm exp}$ term dominates in the
near-field regime, so that the total heat transfer changes from
being independent of separation at large distances towards
$\mathcal{W}_{\rm exp} \propto {1}/{d^{2}}$ at short distances. The
turning point between $\mathcal{W}_{\rm sin}$ and $\mathcal{W}_{\rm
exp}$ occurs when the separation is approximately equal to the peak
wavelength of the black-body curve of the hot half space. Wien's
displacement law predicts that $\lambda_{max}\approx 9$~$\mu$m when
the temperature is 310~K.

\begin{figure}
  \includegraphics[width=.85\columnwidth]{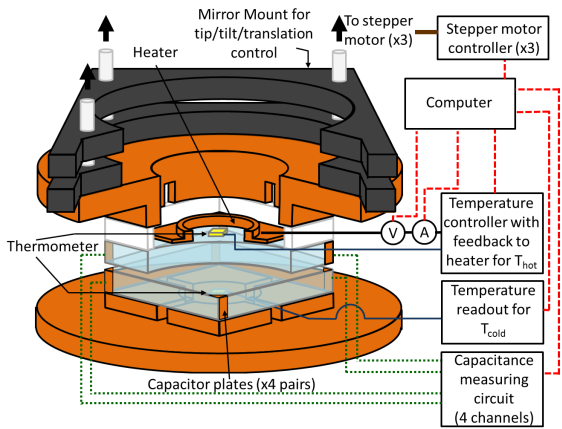}
 \caption{\label{apparatus} Experimental apparatus. Stepper motors
allow adjustment of the spacing, tip, and tilt (read capacitively)
of two sapphire plates. The temperature of the hot plate is
controlled by a feedback circuit, and the power required to maintain
a temperature difference gives the heat transfer from the hot plate
to the cold plate and to the thermal bath.}
\end{figure}

A sketch of our apparatus is shown in Fig.~\ref{apparatus}. It is
designed around two 50$\times$50$\times$5 mm$^3$ sapphire plates.
These have a specified flatness of $\lambda /8$ @ 633 nm per inch on the
largest surfaces and are cut such that the c axis is perpendicular
to these surfaces (z-cut). Sapphire was used because it has good
thermal conductivity. It is also a candidate for the test masses of
future gravitational-wave detectors.

One of the plates, henceforth called the cold plate, is attached to
the thermal bath that is the vacuum chamber. The other plate, the
hot plate, is thermally isolated from the bath by a Macor spacer
attached to the back side of the hot plate. The hot plate also has a
heater wound on a copper ring which is itself attached to the back
of the plate. The heater current and voltage, after correcting for
lead resistance, give the power required to maintain a given
temperature difference between the hot and cold plates.  Both plates
have a Si-diode thermometer fastened to their backs to read the
temperature (and for the hot plate  to control it). Both plates have
all four corners coated with an approximately 200 nm thick layer of
sputtered copper.  These coatings have areas about 1~mm$^2$ and
serve as capacitor plates that are read by four 24-bit
capacitance-to-digital converter circuits\cite{AnalogDevices} to
measure the separation and angular misalignments of the plates. 
The metal film is wrapped around to the sides of
the sapphire to allow electrical contact to the electrodes. 

The cold plate is glued to a copper disk, which in turn is attached
to the experimental structure. The Macor spacer on the back of the
hot plate is attached to a modified kinematic mirror mount which
allows for z-axis linear movement and tip and tilt angular
adjustment by turning the three adjustment screws in the back. Three
stepper motors turn screws on the kinematic mount via gear reduction
boxes; each motor step translates to a linear movement of the hot
plate by 35 nm. The components are held together by an ``L" shaped
backbone (not shown) to give rigidity. The assembly is located in a
UHV chamber, with a base pressure below $2\times 10^{-7}$ Torr,
making gas conduction negligible. Signals to the stepper motors,
capacitance readouts, temperature readouts, and current and voltages
to the heater are all controlled and/or read by a LabVIEW computer
program.

Each pair of capacitor plates is calibrated by taking capacitance
readings as the plates are driven together one step of the stepper
motors at a time. A fit is made to $C=\epsilon_0{a}/{d} + C_{\rm
stray}$ where $\epsilon_0$ is the dielectric constant of the vacuum,
$a$ is the capacitor area, and  $C_{\rm stray}$ is a parallel
contribution independent of separation. The data fits the equation
above very well, with $R^{2}$ values greater than 0.999. The fitted
value of $a$ equals the metalized area within our knowledge of this
area; $C_{\rm stray}\approx 0.4$~pF. The average capacitance gives
the distance while the individual readings are used to correct the
alignment by sending steps to the motors controlling tip and tilt.

To measure the heat transfer coefficient we compute $ \mathcal{W}  =
{P}/[{A(T_{\rm hot} - T_{\rm cold})}]$, where $P$ is the power
dissipated in the heater, $A$ is the plate surface area, and $T_{\rm
hot}$ and $T_{\rm cold}$ are the temperatures of the hot and cold
plates, respectively. The data are a sum of parallel heat pathways,
including thermal conduction through the Macor spacer and the other
parts of the hot-plate holder, radiation to the thermal bath, and
the contributions of the near- and far-field radiation between the
two plates.  We can observe the near-field effect because it is the
only one of these that will change with plate separation. The
thermal conduction and radiation-to-the-bath paths just add a
constant offset to the radiative heat transfer that the model
predicts.

\begin{figure}
\includegraphics[width=.8\columnwidth]{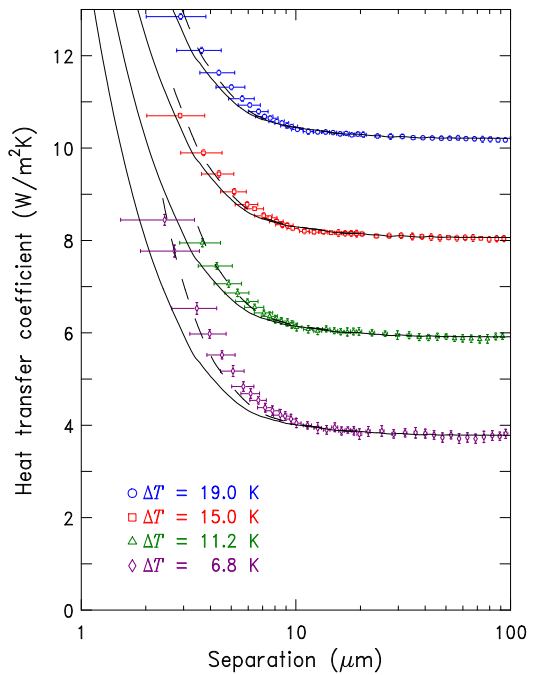}
\caption{\label{Fig:WvsD}
Heat transfer coefficient vs.~distance. The curves are each offset vertically by
2~W/m$^2\cdot$K from the one below. The points are the data, with
error bars, determined from the scatter in the heat transfer
measurements and the uncertainty in the distance calibration. The
solid lines are the theoretical predictions for flat plates while
the dashed lines are the theoretical predictions for slightly curved
plates (see text). Each measured curve has a reproducible addendum due
to other heat leaks which are not included in the model and which
has been subtracted from the data. The temperatures are (top to
bottom): $T_{\rm hot} = 327.0$ K, $T_{\rm cold} = 308.0$ K; $T_{\rm
hot} = 322.0$ K, $T_{\rm cold} = 307.0$ K; $T_{\rm hot} = 317.0$ K,
$T_{\rm cold} = 305.8$ K; $T_{\rm hot} = 312.0$ K, $T_{\rm cold} =
305.2$ K.}
\end{figure}

Data for the heat transfer coefficient  versus
distance were collected for four temperature differences.
(See Fig.~\ref{Fig:WvsD}.)  Each one shows that near-field heat
transfer exists and that the data and model agree reasonably well.
Each run covers a separation range of about 2-100 $\mu$m. The only
freedom in the fit is an offset to the model. All our measurements
have an offset of $0.0435 \pm 0.0004$ W/K, completely consistent
with thermal conduction through the Macor spacer and the rest of the
support for the hot plate, in parallel with radiation from the rear
surface of the plate. The chamber pressure for these runs ranged
between $5 \times 10^{-7}$ Torr and $2 \times 10^{-7}$ Torr. The
vacuum chamber is held at constant temperature of 30.0$\,^{\circ}$C
(303.2~K). The hot plate is brought close to the cold plate step by
step.  Each datum is an average of a set of 500 values, in both heat
transfer coefficient and distance, taken after the system has
reached thermal equilibrium. It takes 30 to 45 minutes after moving
to a new position to reach thermal equilibrium. We calculate the
average and its standard deviation and plot these in
Fig.~\ref{Fig:WvsD}.

Theoretical curves match the experimental data well. However,
although the agreement is within experimental errors, there does
appear to be a systematic offset: the theory predicts a slightly
lower heat transfer coefficient at each separation than we measure.
Alternatively, the plates could be slightly closer than measured by
the capacitive readout. We believe that the latter explanation is
correct. Simulations of the heat transfer between two convex plates
(shown as dashed lines) eliminate the systematic error when the
radius of curvature in each plate is $\sim$1~km,corresponding to
deviations from flatness of 500~nm. Subsequently, we measured the
plate curvatures using optical flats and green mercury light; the
Newton's rings interference pattern gave a central displacement of 
$170 \pm 30$~nm  with respect to the perimeter, a little 
smaller but of the same order of magnitude as the simulation gave.

\begin{figure}
\includegraphics[width=.8\columnwidth]{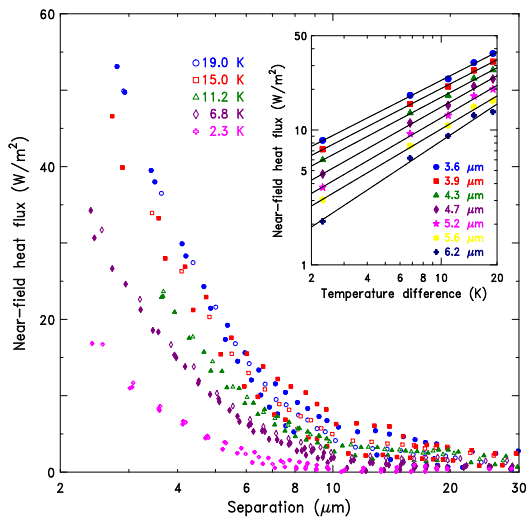}
\caption{\label{Fig:WvsDAll} Near-field heat flux vs.~distance for
multiple temperature differences and multiple runs. (One run is
shown with open symbols and other runs with filled symbols.)  
The inset
shows the near-field heat flux vs.~temperature difference for
several specific separation distances.}
\end{figure}

Figure \ref{Fig:WvsDAll} shows the heat flux caused by
near-field effects. Both far-field heat transfer and the offset due
to heat leaks to the thermal bath have been subtracted. For each
temperature difference, data are shown for several distinct data
runs; these agree very well. The inset shows the dependence on the
temperature difference of the plates. For distance values where the
near-field effects dominate, the heat flux is linear in the
temperature difference.  Both data and model were fitted to
 \begin{equation}
    \label{EQ:WvsT}
    \phi(\Delta T,d) = G(d) \left( \Delta T\right) ^{\alpha (d)}
\end{equation}
where $\phi(\Delta T,d)$ is the total near-field heat flux, 
$G(d)$ is a multiplicative factor, and $\alpha (d)$ is
the exponent for $\Delta T$. Each curve follows a power
law in $\Delta T$,  with the exponent varying from 0.70 at small
distances to 0.91 at larger distances. The differential heat transfer, Eq.~\ref{EQ:W}, has $\alpha (d) = 
1$; the 
finite temperature differences used in the experiment bring in higher-order terms.

In summary we have measured near-field heat transfer  across a small
gap for a parallel-plane geometry.  The data agree quite well with
the theory of Polder and Van Hove\cite{Polder}. The experiments 
demonstrate that significant amounts of heat can be transferred via evanescent
radiation in the near-field regime. 

Our research is supported by the
National Science Foundation through Grant No. PHY-0855313. A.A.A. was
supported by the University of Florida Physics REU Site through NSF
grant DMR-0552726.


\end{document}